\documentclass{astrobul}

\usepackage{graphicx}
\usepackage{natbib}

\usepackage{color}

\begin{document}

\journalinfo{2026}{81}{2}{1}[12]

\title{INFLUENCE OF THE RESONANCE RING GRAVITY
ON THE STELLAR VELOCITY DISTRIBUTION NEAR THE OLR OF THE GALACTIC
BAR}

\author{A.~M. Melnik\address{1}\email{anna@sai.msu.ru},
  E.~N.~Podzolkova\address{1,2},
  \addresstext{1}{Sternberg Astronomical Institute, Lomonosov Moscow
State University, Universitetskij pr. 13, Moscow 119234,  Russia}
  \addresstext{2}{Faculty of Physics, Lomonosov Moscow State University, Leninskie
Gory 1-2, Moscow 119991, Russia} }

\shortauthor{MELNIK and PODZOLKOVA}

\shorttitle{RESONANCE RING GRAVITY}

\submitted{September 19, 2025; in final form, March 10, 2026}

\begin{abstract}
We constructed the 2D model of the Galaxy which initially includes an
analytical bar, bulge, disk and halo. The model disk forms the outer
elliptical resonance rings $R_1$ and $R_2$ located near the outer
Lindblad resonance of the bar (OLR), as well as the inner resonance
ring $r$ located near the corotation radius (CR). As the density of
stars in the elliptical rings increased, we introduced additional
gravitational perturbations created by the rings. The radial
component of gravitational perturbations from the elliptical rings,
$F_R$, at a point with the Galactocentric coordinates ($R$, $\theta$)
was represented as a combination of three polynomials in powers
$R/R_e$ or $R_e/R$, where $R_e$ is the distance to the midline
(middle) of the ring at a given angle $\theta$. The azimuthal
component of the disturbances, $F_T$, was calculated using the force
$F_R$. The difference between the values of the force $F_R$ ($F_T$)
calculated using the numerical differentiation of the potential and
using the analytical representation does not exceed 5.7\% (1.3\%) of
the maximum value of the force $F_R$ generated by the elliptical
rings. In general, the gravity of the elliptical rings has little
effect on the process of adjustment of  epicyclic motions near the
OLR of the bar.

\end{abstract}

\section{1. Introduction}

The existence of the bar in the Galaxy is confirmed by the infrared
observations and velocity distribution data \citep{dwek1995,
dehnen2000, fux2001, benjamin2005, cabrera-lavers2007, gerhard2011,
nesslang2016}.

Bar formation in the galactic disks leads to the appearance of
resonance structures such as the nuclear ($n$), inner ($r$), and
outer ($R_1$ and $R_2$) resonance elliptical rings. The nuclear rings
are forming near the Inner Lindblad Resonance (ILR) of the bar, the
inner rings are near the Corotation Radius (CR), and the outer rings
are near the Outer Lindblad Resonance (OLR) of the bar. Of  two outer
rings, the $R_1$ ring is located slightly closer to the galactic
center  and is elongated perpendicular to the bar, while the $R_2$
ring is located farther from the galactic center and is elongated
parallel to the bar \citep{schwarz1981, buta1991, byrd1994, buta1995,
buta1996, rautiainen1999, rautiainen2000, melnikrautiainen2009,
rautiainen2010, melnik2019}. The resonance rings are supported by
stable periodic orbits, near which there are a lot of quasi-periodic
orbits \citep{contopoulos1980, contopoulos1989}.

The kinematical and spatial distributions of young objects (OB
associations, classical Cepheids, star-gas complexes) suggest the
presence of a double outer ring $R_1R_2$ in the Galaxy
\citep{melnikrautiainen2009, rautiainen2010, melnik2011, melnik2015,
melnik2016, melnik2019}.

Previously, we built the distribution of the radial, $V_R$, and
azimuthal, $V_T$, velocities along the Galactocentric distance, $R$,
for a large number of old disk-stars and showed that the values of
the angular velocity of the bar's rotation of $\Omega_b=55$ km
s$^{-1}$ kpc$^{-1}$ and the Sun's position angle relative to the
bar's major axis of $\theta_\odot=-45^\circ$ provide the best
agreement between the models and observations \citep{melnik2021}.

A study of the distribution of stars in the model disk revealed
periodic changes in the morphology of the resonance rings, namely, a
strengthening either leading or trailing segments of the elliptical
rings \citep{melnik2023}. Moreover, the  profiles of the distribution
of the radial velocities, $V_R$, along the distance $R$ demonstrate
the periodic appearance of the humps. Although the height of the
humps is only $1.76\pm0.15$ km s$^{-1}$, their statistical
significance (ratio of value to its error) exceeds $11\sigma$. The
analysis of the orientation of orbits lying near the OLR of the bar
showed that the change in the morphology of the outer rings and the
appearance of the humps are caused by one reason  --- librating
orbits, i.~e. orbits that change their orientation relative to the
major axis of the bar in a limited range of angles
\citep{melnik2024}.

The existence of librating orbits near the Lindblad resonances of the
bar was predicted by Weinberg \citep{weinberg1994}. Theoretical
aspects of the formation  of librating orbits are also given in other
papers \citep{monari2016, monari2017}.

\citet{kondrat'ev2007} obtained formulas for the gravitational
potential of the elementary flat elliptical rings. The main feature
of the flat rings is that they create gravitational disturbances not
only in the outer region, but also in the inner region of the ring,
up to the center of the system, in contract to the three-dimensional
homogeneous elliptical shells. The problem of the potential of flat
rings is closely related to the problem of the potential of thin
disks \citep{letelier2007}.

The goal of this work is to estimate the influence of gravitational
disturbances created by formed resonance rings on the distribution of
the velocities $V_R$ and $V_T$ near the OLR bar and on the process of
the appearance of librating orbits.

We will show that the gravity created by the elliptical rings has
little effect on the process of tuning epicyclic motions near the OLR
of the bar.

\section{2. Observations}

We use data of the {\it Gaia} DR3 catalog  \citep{prusti2016,
katz2018, brown2021, lindegren2021, vallenari2023} to build the
observational distributions of the radial, $V_R$, and azimuthal,
$V_T$, velocities along the Galactocentric distance, $R$. Our sample
contains $\sim 9.7\times10^6$ stars lying near the Galactic plane,
$|z|<0.2$ kpc, and in the sector of the azimuthal angles
$|\theta|<15^\circ$, having the parallax to parallax error ratio
$\varpi/\varepsilon_\varpi>5$, the error $\textrm{RUWE}<1.4$, and the
line-of-sight velocity $V_r$ measured by the {\it Gaia} spectrometer.
The median values of the radial and azimuthal velocities of stars
were calculated in the $\Delta R=0.25$-kpc wide bins. Random errors
in the determination of the median velocities $V_R$ and $V_T$ in the
distance bins are within 0.02--0.6 km s$^{-1}$ and strongly depend on
the number of stars in a given bin, but for the  distance range
$R=6$--9 kpc they do not exceed 0.1 km s$^{-1}$ \citep[more
see][]{melnik2021}.

The distance of the Sun to the Galactic center is taken to be
$R_0=7.5$ kpc \citep[][]{glushkova1998, nikiforov2004,
eisenhauer2005, bica2006, nishiyama2006, feast2008, groenewegen2008,
reid2009b, dambis2013, francis2014, boehle2016, branham2017,
iwanek2023}. The choice of the $R_0$ value in the range of 7--9 kpc
has practically no effect on our results.

The angular velocity of the rotation of the Galactic disk  at the
solar distance is taken to be $\Omega_0=30$ km s$^{-1}$ kpc$^{-1}$,
which is consistent with the kinematics of the OB associations
\citep{melnik2020}. With this choice of $R_0$ and $\Omega_0$, the
azimuthal velocity of the disk rotation at the solar distance is
$V_T=225$ km s$^{-1}$.

\section{3. Models}

We consider 2D models of the Galaxy including the analytical Ferrers
bar \citep{freeman1972, athanassoula1983, pfenniger1984,
sellwood1993, binney2008}, an exponential disk, a classical bulge and
halo. The masses of the disk, bar and bulge are $3.25 \times
10^{10}$, $1.2 \times 10^{10}$ and $5 \times 10^{9}$ M$_\odot$,
respectively. The characteristic horizontal scale of the exponential
disk is 2.5 kpc and it does not change during the simulation. The
rotation curve of the model disk is flat at the periphery and
corresponds to an angular velocity of rotation of $\Omega_0=30$ km
s$^{-1}$ kpc$^{-1}$ at the distance of the Sun, which is consistent
with the observations \citep{melnik2020}.

In this paper, we use three models of the Galaxy. Model 1 does not
take into account the gravity from the elliptical rings and is used
in our previous works \citep{melnik2021, melnik2023, melnik2024,
melnik2025}. Model 2 includes gravitational forces from three
elliptical rings ($R_1$, $R_2$ and $r$). Model 3 takes into account
gravity from only  two outer elliptical rings ($R_1$ and $R_2$). In
calculation of the gravitational forces from the elliptical rings, we
use the analytical expressions obtained in Section 4.2.

In our models, the major and minor semi-axes of the bar are $a=3.5$
and $b=1.25$ kpc, respectively. The angular velocity of the bar is
taken to be $\Omega_b=55$ km s$^{-1}$ kpc$^{-1}$ and does not change
during the simulation. The positions of the  CR and OLR of the bar
correspond to the distances of $R_{CR}=4.04$ and $R_{OLR}=7.00$ kpc.
The bar turns on gradually gaining its full strength over a time of
$T_g=0.45$ Gyr. The angle $\theta$ is measured from the direction of
the major axis of the bar counterclockwise and increases in the sense
of the Galactic rotation, and the angle $\theta_\odot$ determines the
position of the Sun relative to the major axis of the bar and is
adopted to be $\theta_\odot=-45^\circ$. Since our models have the
order of symmetry $m=2$, both values of the position angle,
$\theta_\odot=-45^\circ$ and $135^\circ$, are equivalent.

The gravitational forces generated by the elliptical rings are turned
on gradually and reach their full strength during the  time interval
of  $T_2=0.5$ Gyr. The time of the beginning of the growth of the
rings is discussed in Section 4.1 and is taken to be $T_1=0.5$ Gyr.
Thus, the gravitational forces from the elliptical rings reach its
final values by the time $t=T_1+T_2=1.0$ Gyr and remain unchanged
thereafter. Since the orientation of the elliptical rings is
determined by the bar, the rings rotate with the angular velocity of
the bar. In this work, we consider only elliptical rings whose
morphology does not change with time.

The mass of the disk decreases as elliptical rings grow, so that the
total mass of the new exponential disk and the rings is conserved.

The models contain $2 \times 10^{6}$ massless particles. The initial
radial dispersion of stars of the model disk at the distance of the
Sun is $\sigma_R \approx 30$ km s$^{-1}$, i.~e. we consider the
motions of old stars of the disk. The simulation step is 0.01 Myr.
The simulation time is 6 Gyr. We choose such a long simulation time
to show that the gravity of the elliptical rings has little effect on
the formation of the humps on the profiles of the distribution of the
velocity $V_R$ along the distance $R$.

\section{4. RESULTS}
\subsection{4.1 Density distribution in the model disk}

\begin{figure*}
\resizebox{\hsize}{!}{\includegraphics{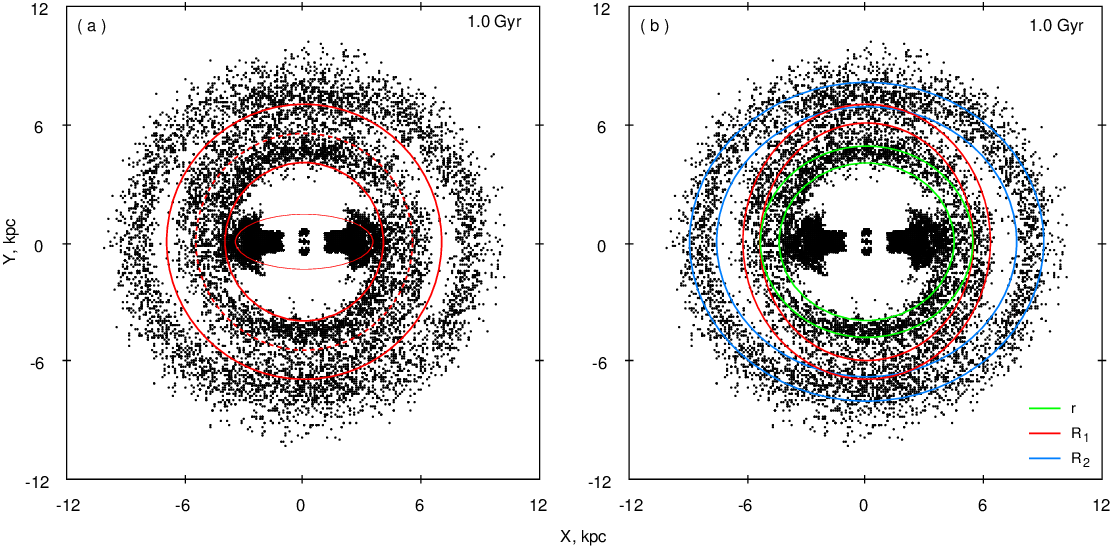}} \caption{ (a)
Distribution of stars in the Galactic plane in model 1 at the time
instant $t=1.0$ Gyr processed by the program that increases the
contrast. The parameter $h$, which controls the contrast, is $h=1.5$.
Only 10\% of  particles are shown. The positions of the bar
(ellipse), the circles of radii CR and OLR (solid red lines) and the
$-4/1$ resonance (dashed red line) are also shown. The Galaxy rotates
counterclockwise. The $X$ and $Y$ axes lie in the Galactic plane and
are parallel to the major and minor axes of the bar, respectively. We
can clearly see the presence of the inner ring $r$ located outside
the circle of the CR. The outer ring $R_1$ is represented by two
trailing spiral arms located in the region between the resonances
$-4/1$ and OLR. The outer ring $R_2$ is elongated parallel to the bar
and lies outside the OLR radius, $R>R_{OLR}$. (b) The conditional
boundaries of the rings are shown by the green ($r$), red ($R_1$) and
blue ($R_2$) ellipses, respectively.} \label{distrib}
\end{figure*}

We tried to determine the positions of the elliptical rings in the
model disk. Figure~\ref{distrib}a shows the distribution of stars in
the Galactic plane in model 1, which does not take into account the
gravity from the elliptical rings, at the time moment $t=1.0$ Gyr
from the start of  simulation processed by the program that enhances
the contrast. For each star, we calculated the density of stars
located within a radius of 50 pc from it, $\Sigma_1$, which  then is
compared with the initial density $\Sigma_0$ at the time instant
$t=0$, when the distribution of stars was purely exponential. If
$\Sigma_1/\Sigma_0\le 1.0$, then the star is not included in the
image, if $\Sigma_1/\Sigma_0\ge h$, then the star is included in the
image with a probability of P=100\%. If the density ratio has an
intermediate value, $1.0<\Sigma_1/\Sigma_0<h$, then the star is
included in the image with the probability of:

\begin{equation}
P=\frac{\Sigma_1-\Sigma_0}{\Sigma_0 (h-1)}\ 100\%,
 \label{p}
\end{equation}

\noindent which linearly changes from 0 to 100\% with increasing the
ratio $\Sigma_1/\Sigma_0$ from 1.0 to $h$. The value of the parameter
$h$ was adopted to be $h=1.5$. \citep[see also Section 4.1
in][]{melnik2023}.

Figure~\ref{distrib}a also shows the positions of the bar (ellipse),
the circles of radii CR and OLR (solid red lines) and the resonance
$-4/1$ (dashed red line) in the model disk. The Galaxy rotates
counterclockwise. We can clearly see the presence of the inner ring,
$r$, located outside the circle of the CR, and the outer ring $R_2$,
elongated parallel to the bar and located outside the circle of the
OLR radius, $R>R_{OLR}$. We have chosen for illustration the time
moment $t=1.0$ Gyr, when the density of the ring $R_2$ has  maximum
value.

The outer ring $R_1$ at the time moment  $t=1.0$ Gyr is represented
by two trailing spiral arms located in the region between the CR and
OLR of the bar. If we  chosen for illustration the moment $t=1.5$
Gyr, then the ring $R_1$ would be represented by two leading spiral
arms, which is a consequence of the periodic change in the morphology
of the outer rings \citep{melnik2023}.

Figure~\ref{distrib}b shows the conditional boundaries of the
elliptical rings. The inner ring, $r$, and the outer rings, $R_1$ and
$R_2$, are outlined by green, red and blue ellipses, respectively.
The boundaries of the rings were chosen in a such way that the
density of stars in the rings has maximum value after averaging over
the time interval of 0--6 Gyr. The boundaries and the midline
(middle) of each ring are similar ellipses, i.~e. they have the same
ratio of their semi-axes.

Table 1 lists the values of the major and minor semi-axes of the
ellipses, $a_r$ and $b_r$, which outlined  the positions of the
midlines of the rings, the half-width of the rings in the direction
of the major axis, $\Delta a$, the angle  $\theta_0$, which
determines the orientation of the ring with respect to the bar's
major axis, the average density, $\Delta \overline{\Sigma}$, and the
mass of the rings, $M$.

\begin{table*}
 \caption{\normalsize Elliptical rings}
 \centering
\begin{tabular}{lcccccc}
\\[-7pt]\hline\\[-7pt]
 Type of a ring & $a_r$, kpc & $b_r$, kpc & $\Delta a$, kpc & Orientation & $\Delta \overline{\Sigma}$, $M_\odot$ kpc$^{-2}$ &  $M$, $M_\odot$ \\
\\[-7pt]\hline\\[-7pt]
Inner ring $r$ & 4.9 & 4.5 & 0.5 & $\theta_0=0^\circ$ & $5.12\times 10^6$ & $144.73\times 10^6$ \\
\\[-7pt]\hline\\[-7pt]
Outer ring $R_1$ & 6.5 & 6.0 & 0.5 & $\theta_0=90^\circ$ &  $0.49\times 10^6$ & $18.38\times 10^6$ \\
\\[-7pt]\hline\\[-7pt]
Outer ring $R_2$ & 8.3 & 7.7 & 0.7 &  $\theta_0=0^\circ$ & $1.14\times 10^6$  & $77.05\times 10^6$ \\
\\[-7pt]\hline\\[-7pt]
\end{tabular}
\label{ellipse}
\end{table*}

\begin{figure*}
\resizebox{\hsize}{!}{\includegraphics{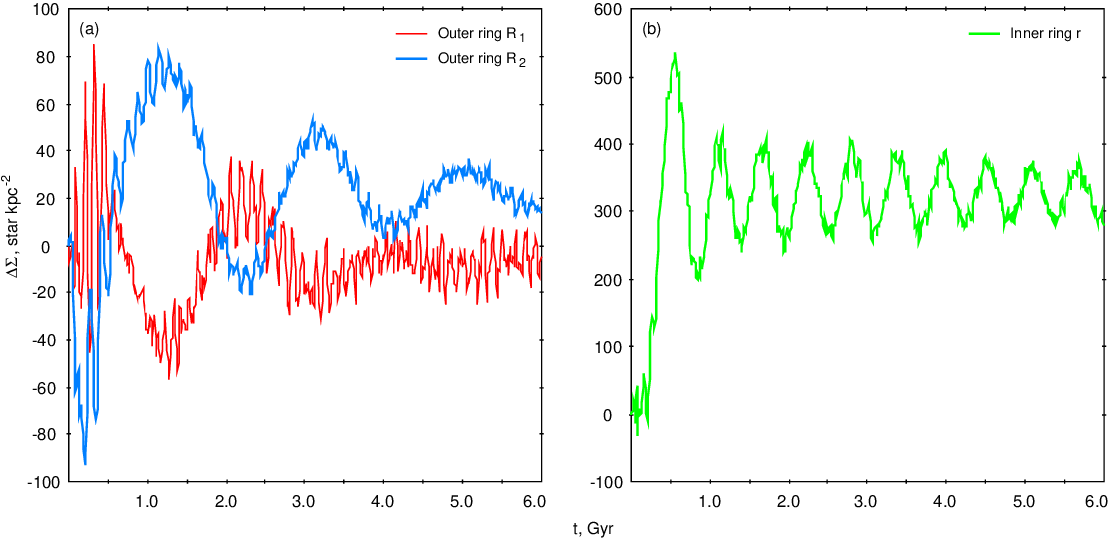}} \caption{Variations
in the average density of stars in the outer rings $R_1$ and $R_2$
(a) and in the inner ring $r$ (b). The average density at different
time moments is calculated relative to the average density at $t=0$,
when the distribution of stars in the model disk is exponential:
$\Delta \Sigma(t)=\Sigma(t)-\Sigma(0)$. Variations in the density of
stars in the rings $R_1$ and $R_2$ occur with a period of
$P=1.9\pm0.1$ Gyr and practically in antiphase. The period of density
oscillations in the inner ring $r$ is $P=0.58\pm0.01$ Gyr.}
\label{oscil}
\end{figure*}

The problem is that the average density of stars in the rings changes
considerably with time. Figure~\ref{oscil} shows the oscillations of
the average density of stars in the $r$, $R_1$, and $R_2$ rings. We
consider the oscillations of the average density relative to its
value at the time moment $t=0$, when the distribution of stars in the
model disk is exponential: $\Delta \Sigma=\Sigma(t)-\Sigma(0)$. It is
seen that oscillations include fast and  slow components.

Figure~\ref{oscil}a shows that slow oscillations of the average
density of stars, $\Delta \Sigma$, in the $R_1$ and $R_2$ rings have
nearly equal periods and occur practically in antiphase. In addition,
the amplitude of slow oscillations decay with time. The periods of
slow oscillations of the $\Delta \Sigma$ in the $R_1$ and $R_2$ rings
are $P=1.93\pm0.03$ and $P=1.98\pm0.02$ Gyr, respectively, which is
close to the median period of oscillations of the librating orbits,
$P \approx 2.0$ Gyr \citep{melnik2024}. The fact that slow
oscillations of the density in the $R_1$ and $R_2$ rings occur in
antiphase is a consequence of the circumstance  that the librating
orbits near the OLR periodically change their average radius,
$\overline{R}$. Since the $R_1$ ring is located closer to the center
of the Galaxy than the $R_2$ ring, the probability that a star will
fall into the region of the $R_2$ ring ($R_1$) increases (decreases)
with increasing average distance $\overline{R}$ \citep[Fig.~11e
in][]{melnik2023}.

To estimate the mass of the $R_1$ and $R_2$ rings, we  use  maximum
value of slow oscillations of the density, $\Delta \Sigma$, after the
bar reaches its full power, $t>0.45$ Gyr. At this time interval
$\Delta \overline{\Sigma}$ reaches values of 30 ($R_1$) and 70
($R_2$) particle kpc$^{-2}$. Further we neglect the variations in the
ring density with time and take the values of $\Delta
\overline{\Sigma}$ for the outer rings to be 30 ($R_1$) and 70
($R_2$) particle kpc$^{-2}$, which, with the disk mass of
$M_d=3.25\times 10^{10}$ M$_\odot$ and the number of particles $N= 2
\times 10^6$, corresponds to the densities of $0.49\times10^6$ and
$1.14\times 10^6$ M$_\odot$ kpc$^{-2}$, respectively (Table 1).

Fig.~\ref{oscil}b shows that the density perturbation, $\Delta
\Sigma$, in the inner ring $r$ reaches a maximum and then begins to
oscillate near the average value, and  the amplitude of slow
oscillations decreases with time. The position of the maximum $\Delta
\Sigma$ corresponds to the time moment $t=0.53$ Gyr, which is close
to the moment when the bar reaches its full power, $T_g=0.45$ Gyr.
The oscillations of $\Delta \Sigma$ occur near the average value of
$\Delta \Sigma=315\pm3$ particle kpc$^{-2}$  with the period of
$P=0.58\pm0.01$ Gyr. This period is close to the period of
oscillations of stars in long-period orbits near the equilibrium
points $L_4$ and $L_5$, which  is $0.565\pm0.002$ Gyr in our model.
Hereinafter,  we will use the value of the average density of the
inner ring of $\Delta \overline{\Sigma}=315$ star kpc$^{-2}$, which
in solar mass units is $5.12\times 10^6$ M$_\odot$ kpc$^{-2}$ (Table
1).

Note that  $\Delta \Sigma$ characterizes the perturbation of the
average density of stars inside the area of the ring and changes with
time (Fig.~\ref{oscil}), but  $\Delta \overline{\Sigma}$ is an
estimate of the average density of stars in the ring without taking
into account changes over time (Table 1).

Using estimates of the average ring density $\Delta
\overline{\Sigma}$ and the formula for the area of an ellipse ($S=\pi
a b$), it is easy to show that the time-averaged masses of the inner
ring, $r$, and the outer rings, $R_1$ and $R_2$, are $144.73 \times
10^6$, $18.38 \times 10^6$ and $77.05 \times 10^6$ M$_\odot$,
respectively (Table 1).

In simulations, we neglected  oscillations in the densities and
masses of the rings. In general, such a simplification is acceptable,
since we are interested in the upper estimate of the velocity
perturbations caused by the gravity from the elliptical rings.

\subsection{4.2 Gravitational forces created by an
elliptical ring. Analytical representation}

We  consider the gravitational potential of the Galactic disk to be a
superposition of the potentials of the exponential disk and the
potentials of the elliptical rings.

Knowing the density distribution in an elliptical ring, we can
calculate the gravitational potential at an arbitrary point with the
coordinates ($R$, $\theta$):

\begin{equation}
\Phi(R,\theta)= -G \int_{0}^{2\pi}\int_{R_e-\Delta R}^{R_e+\Delta R}  \\
 \label{phi}
\end{equation}
$$
\;   \frac{\Delta
\Sigma(R',\theta')}{\sqrt{R'^2+R^2+2RR'\cos(\theta'-\theta)}}\, R'
dR' d\theta'
$$

\noindent \citep[e.g.,][]{kalnajs1971, binney2008}. The
Galactocentric angles $\theta$ and $\theta'$ are measured from the
major axis of the bar in the direction of the Galactic rotation. The
distance from the Galactic center to the midline of the elliptical
ring, $R_e$, is determined by the following expression:

\begin{equation}
R_e=\frac{b_r}{\sqrt{1-e^2\cos^2(\theta-\theta_0)}},
 \label{R_e}
\end{equation}

\noindent where $\theta_0$ is the angle between the  major axis of
the bar and the  major axis of the elliptical ring, and $e$ is the
eccentricity of the elliptical ring:

\begin{equation}
e^2=1-\frac{b_r^2}{a_r^2},
 \label{e}
\end{equation}

\noindent where $a_r$ and $b_r$ are the major and minor semi-axes of
the ellipse defining the position of the midline of the ring,
respectively. Since the boundaries of the ring are similar ellipses,
the width of the ring, $\Delta R$, in a given direction is determined
by the relation:

\begin{equation}
\Delta R=R_e \Delta a/a_r,
 \label{delta_R}
\end{equation}

\noindent where $\Delta a$ is the half-width of the ring in the
direction of the major axis of the ellipse.

We assume that the density distribution across the elliptical ring
varies in the cosine law:

\begin{equation}
\Delta \Sigma= \Sigma_m \, \cos \gamma,
 \label{cos}
\end{equation}

\noindent where the value of $\gamma$ is determined from the
expression:

\begin{equation}
\gamma=\frac{\pi}{2\Delta R}\,(R-R_e)
 \label{gamma}
\end{equation}

\noindent and takes the values of $-\pi/2$ and $\pi/2$ on the inner
and outer boundaries of the ring, respectively. On the midline of the
ring $\gamma=0$, and the density reaches its maximum value,
$\Sigma_m$. The cosine law in the density distribution across the
ring avoids density jumps at the ring's boundaries.

It is easy to show that the mass of the elliptical ring, $M$, is:

\begin{equation}
M=8 b_r \Delta a \Sigma_m,
 \label{phi}
\end{equation}

\noindent and maximum density of stars on the midline of the ring,
$\Sigma_{m}$, must be related to the average density of the ring,
$\overline{\Sigma}$, as follows:

\begin{equation}
\Delta \Sigma_{m}=\frac{\pi}{2} \, \Delta \overline{\Sigma}.
 \label{s}
\end{equation}

The number of integration intervals $N_{int}$ is chosen to be
$N_{int}=2000$, which corresponds to the integration steps in
$\theta'$ and $R'$ equal to $d \theta'=2\pi/N_{int}=0.18^\circ$ and
$dR'=2\Delta R/N_{int}=0.5$--0.7 pc. The potential distribution is
presented in the form of a table $\Phi(R_i,\theta_j)$ with steps in
$R$ and $\theta$  equal to $dR=0.1$ kpc and $d\theta=10^\circ$,
respectively.

Knowing the potential distribution, we can calculate the  radial and
azimuthal components of the gravitational force acting on a particle
of unit mass from the elliptical ring:

\begin{equation}
F_R(i,j)=\frac{\Phi(i+1,j)-\Phi(i-1,j)}{2dR},
 \label{d_fr}
\end{equation}
\begin{equation}
F_T(i,j)=\frac{1}{R}\frac{\Phi(i,j+1)-\Phi(i,j-1)}{2d\theta},
 \label{d_ft}
\end{equation}

\noindent and obtain the distribution of gravitational disturbances
created by the elliptical ring in the form of a table of size (110,
36) along $R$ and $\theta$.

\begin{figure*}
\centering \resizebox{\hsize}{!}{\includegraphics{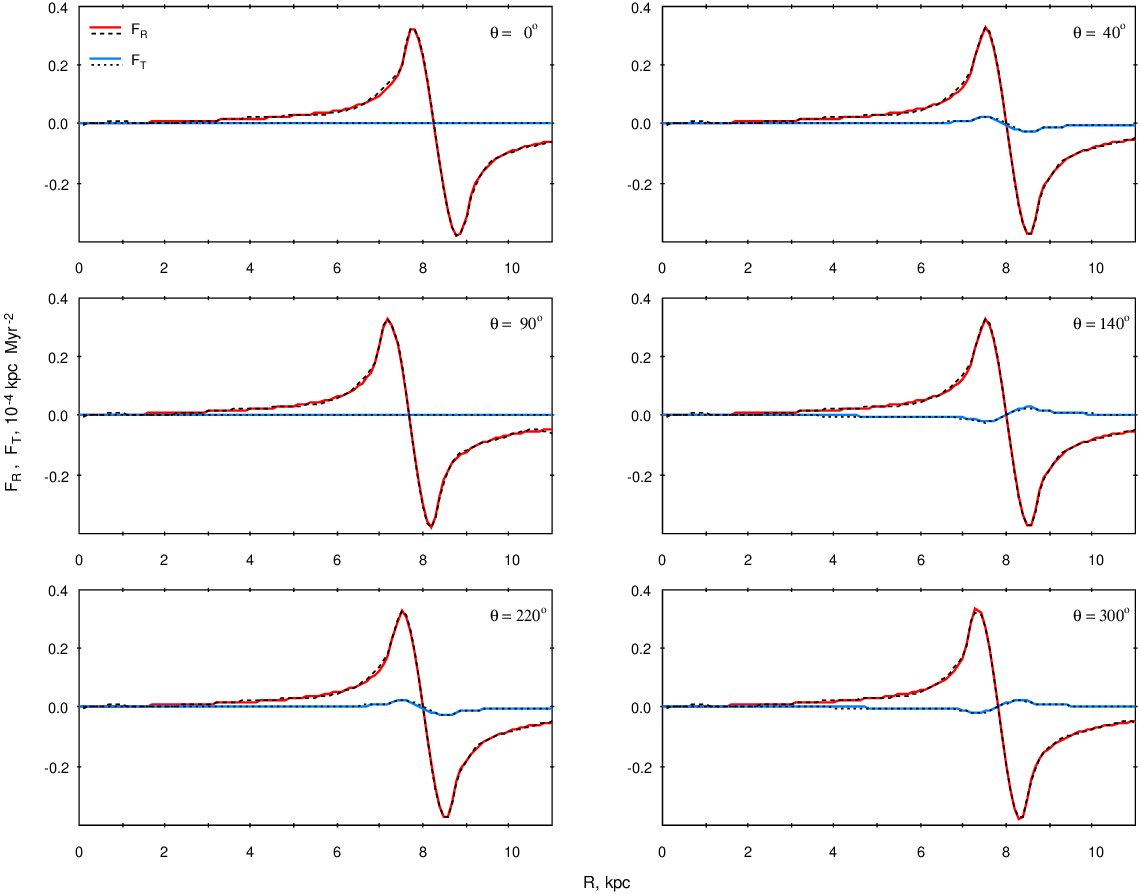}}
\caption{ Distributions of the radial, $F_R$, and azimuthal, $F_T$,
components of the gravitational forces acting on a particle of unit
mass from the elliptical ring $R_2$. The variations of the forces
along the Galactocentric distance $R$ are shown for different values
of the Galactocentric angle $\theta$. The forces $F_R$ and $F_T$
obtained by numerical differentiation of the potential
(Eq.~\ref{d_fr} and \ref{d_ft}) are shown in red and blue,
respectively, and the values calculated by formulas
~\ref{fr_a}--\ref{ft_d} are shown as black dotted lines. We can see a
good agreement between the numerical and analytical values of the
forces $F_R$ and $F_T$.} \label{force}
\end{figure*}

Figure~\ref{force} shows the distributions of the radial $F_R$ and
azimuthal $F_T$ components of the gravitational forces acting on a
particle of  unit mass from the elliptical ring $R_2$. The changes in
the forces along the distance $R$ are shown for different values of
the angle $\theta$. Since the ring $R_2$ is elongated along the bar,
i.~e. the angle $\theta_0=0^\circ$ (Eq.~\ref{R_e}, Table 1), the
directions $\theta=0$ and $90^\circ$ correspond to the major
($\theta=0^\circ$) and minor ($\theta=90^\circ$) axes of the bar. The
forces $F_R$ and $F_T$ obtained by numerical differentiation of the
potential (Eq.~\ref{d_fr} and \ref{d_ft}) are shown by the red and
blue lines, respectively, while the forces calculated by the
analytical  representation (see below) --- by the dashed lines.

The force $F_R$ is always directed toward the midline of the ring: in
the inner region of the ring it has positive values (directed away
from the center of the Galaxy), and in the outer region it has
negative values (directed toward the center of the Galaxy). At the
midline of the ring itself, the force $F_R$ is zero
(Fig.~\ref{force}). Note that the gravitational forces inside the
inner boundary of a flat elliptical ring are not  zero, which is a
general property of flat rings \citep[][]{kondrat'ev2007}.

The positions of the inner and outer boundaries of the elliptical
ring, $R_{in}$ and $R_{out}$, are determined by the relations:

\begin{equation}
R_{in}=\frac{a_r-\Delta a}{a_r} R_e,
 \label{R_in}
\end{equation}
\begin{equation}
R_{out}=\frac{a_r+\Delta a}{a_r} R_e.
 \label{R_out}
\end{equation}

We assume  the distribution of the force $F_R$ over the distance $R$
and the angle $\theta$ to be a combination of three polynomials in
powers of $R$. If the point under consideration lies inside the inner
boundary of the ring, i.~e. $0<R<R_{in}$, then $F_R$ can be
represented as a fifth-order polynomial in powers of $p=R/R_e$:

\begin{equation}
F_R=M \,(a_0+a_1p +a_2p^2+a_3p^3+a_4p^4+a_5p^5), \label{fr_a}
\end{equation}

\noindent where $M$ is the mass of the ring. Note that in this region
the distance $R$, and hence $p$, can vanish. If the point lies inside
the elliptic ring, i.~e.  $R_{in}\le R \le R_{out}$, then $F_R$ can
be calculated using a fourth-order polynomial in powers of $s=R_e/R$:

\begin{equation}
F_R=M \,(b_0+b_1s +b_2s^2+b_3s^3+b_4s^4). \label{fr_b}
\end{equation}

\noindent And finally, if the point lies outside the outer boundary
of the ring, $R > R_{out}$, then $F_R$ can be calculated using
another fourth-order polynomial in powers of $s=R_e/R$:

\begin{equation}
F_R=M \, (c_0+c_1s +c_2s^2+c_3s^3+c_4s^4). \label{fr_c}
\end{equation}

\noindent The degree of a polynomial corresponds to the minimum value
above which  a further increase in the degree does not result in a
significant increase in the approximation accuracy.

Figure~\ref{force} shows that the positions of the extrema in the
distributions of forces $F_R$ and $F_T$ practically coincide.
Furthermore, the profiles of the $F_R$- and $F_T$-distributions
intersect at the points with ordinates close to zero. This means that
the force $F_T$ can be approximately calculated from the force $F_R$
by multiplying  a coefficient that depends only on the angle
$\theta$, or more precisely, on  $2\theta$, since the order of
symmetry of the models is $m=2$:

\begin{equation}
F_T=F_R\,e^2 (d_1 \sin 2\theta+d_2 \sin 4\theta+d_3 \sin 6\theta+d_4
\sin 8\theta). \label{ft_d}
\end{equation}

\noindent We put out of brackets $M$ (Eq.~\ref{fr_a}--\ref{fr_c}) and
$e^2$ (Eq.~\ref{ft_d}) to avoid recalculation of the coefficients for
a small change in the mass of the ring or in its eccentricity.

The parameters $a_0$, .., $a_5$; $b_0$, .., $b_4$; $c_0$, .., $c_4$
and $d_1$, .., $d_4$, calculated for the rings $R_2$, $R_1$ and $r$,
are given in Table~\ref{tab_abcd}. Parameters $d_1$, .., $d_4$ are
dimensionless, but the rest have the dimension [acceleration
mass$^{-1}$], which in the adopted units of measurement corresponds
to the dimension [kpc Myr$^{-2}$ 10$^{-11}$ $M_\odot^{-1}$]. It is
clearly seen that the parameters $a_0$, .., $a_5$; $b_0$, .., $b_4$
and $c_0$, .., $c_4$, calculated for the three rings, differ
significantly, which is probably due to their different positions
relative to the center of the Galaxy. On the other hand, the
parameters $d_1$, .., $d_4$, connected  the forces $F_R$ and $F_T$ in
the rings $R_2$ and $R_1$, coincide in absolute value within the
errors, and the difference in signs is caused by the difference in
their orientation.

For the ring $R_2$, the forces $F_R$ calculated by numerical
differentiation (Eq.~\ref{d_fr}) and by analytical representation
(Eq.~\ref{fr_a}--\ref{fr_c}) differ by no more than 4.3\% from
maximum  force $F_R$ in the ring, while the forces $F_T$, calculated
by two different methods  (Eq.~\ref{d_ft} and \ref{ft_d}), differ by
no more than 1.0\% from the maximum  force $F_R$. Note that  maximum
force $F_R$ in the ring $R_2$ ($\sim 0.4 \times 10^{-4}$ kpc
Myr$^{-2}$) is only 0.7\% of the average value of the total radial
force at the same radius. For the rings $R_1$ and $r$, the similar
estimates are 5.5, 1.2, and 0.2\% and 5.7, 1.3, and 1.7\%,
respectively.

\begin{table*}  \centering  \caption{\normalsize Parameters obtained for the analytical representation of the forces
$F_R$ and $F_T$}
\begin{tabular}{crrr}
\\[-7pt]\hline\\[-7pt]
Parameters & Outer ring $R_2$ & Outer ring $R_1$ &  Inner ring $r$ \\
kpc Myr$^{-2}$ 10$^{-11}$ $M_\odot^{-1}$ &  &  &   \\
\\[-7pt]\hline\\[-7pt]
$a_0$ & $        -0.0008 \pm    0.0001$ & $        -0.0014 \pm    0.0001$ & $        -0.0016 \pm    0.0002$  \\
$a_1$ & $         0.0409 \pm    0.0017$ & $         0.0710 \pm    0.0034$ & $         0.0902 \pm    0.0047$  \\
$a_2$ & $        -0.3555 \pm    0.0118$ & $        -0.6216 \pm    0.0236$ & $        -0.7981 \pm    0.0336$  \\
$a_3$ & $         1.2503 \pm    0.0333$ & $         2.1816 \pm    0.0661$ & $         2.9108 \pm    0.0971$  \\
$a_4$ & $        -1.8171 \pm    0.0404$ & $        -3.1584 \pm    0.0800$ & $        -4.3582 \pm    0.1212$  \\
$a_5$ & $         0.9426 \pm    0.0177$ & $         1.6290 \pm    0.0348$ & $         2.3288 \pm    0.0543$  \\
\\[-7pt]\hline\\[-7pt]
$b_0$ & $          319.5 \pm       5.8$ & $          962.4 \pm      18.0$ & $          484.4 \pm       6.4$  \\
$b_1$ & $        -1194.6 \pm      22.8$ & $        -3652.9 \pm      70.9$ & $        -1826.7 \pm      25.0$  \\
$b_2$ & $         1662.9 \pm      33.9$ & $         5175.6 \pm     104.9$ & $         2565.5 \pm      36.8$  \\
$b_3$ & $        -1021.1 \pm      22.3$ & $        -3244.2 \pm      68.9$ & $        -1590.6 \pm      24.0$  \\
$b_4$ & $          233.3 \pm       5.5$ & $          759.1 \pm      16.9$ & $          367.4 \pm       5.9$  \\
\\[-7pt]\hline\\[-7pt]
$c_0$ & $         -30.91 \pm      0.67$ & $          -5.37 \pm      0.11$ & $          -1.25 \pm      0.02$  \\
$c_1$ & $         157.11 \pm      3.29$ & $          31.21 \pm      0.60$ & $           8.61 \pm      0.16$  \\
$c_2$ & $        -299.10 \pm      6.06$ & $         -67.59 \pm      1.24$ & $         -21.90 \pm      0.37$  \\
$c_3$ & $         252.79 \pm      4.95$ & $          64.64 \pm      1.12$ & $          24.36 \pm      0.38$  \\
$c_4$ & $         -80.06 \pm      1.51$ & $         -23.07 \pm      0.38$ & $         -10.06 \pm      0.15$  \\
\\[-7pt]\hline\\[-7pt]
Parameters, dimensionless& & &  \\
$d_1$ & $         0.5063 \pm    0.0010$ & $        -0.5064 \pm    0.0011$ & $         0.5018 \pm    0.0013$  \\
$d_2$ & $         0.0155 \pm    0.0010$ & $         0.0161 \pm    0.0011$ & $         0.0165 \pm    0.0013$  \\
$d_3$ & $         0.0026 \pm    0.0010$ & $        -0.0025 \pm    0.0011$ & $         0.0023 \pm    0.0013$  \\
$d_4$ & $         0.0006 \pm    0.0010$ & $         0.0004 \pm    0.0011$ & $         0.0003 \pm    0.0013$  \\
\\[-7pt]\hline\\[-7pt]
\end{tabular}
\label{tab_abcd}
\end{table*}

\begin{figure*}
\resizebox{16.0 cm}{!}{\includegraphics{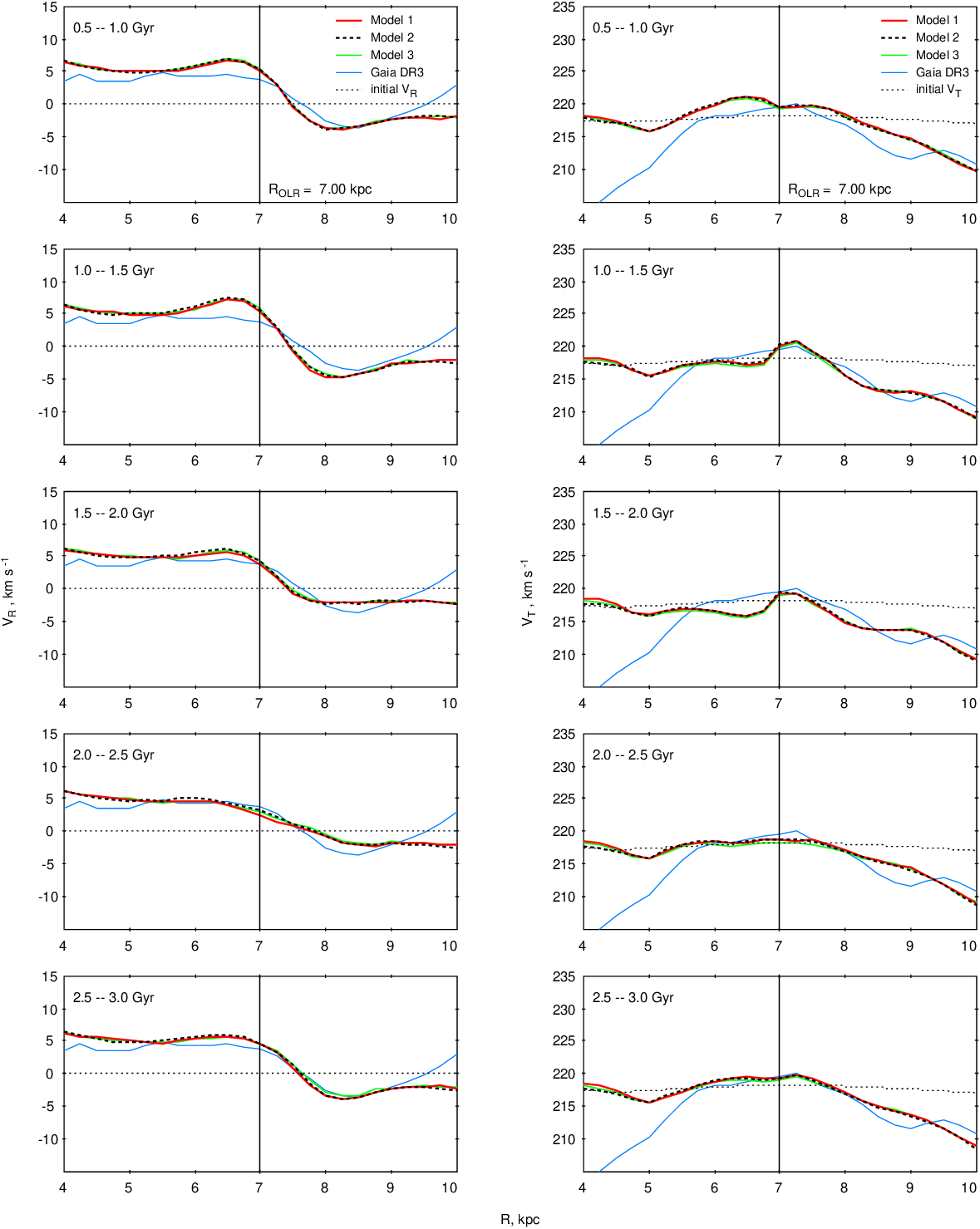}}
\caption{Distributions of the median velocities $V_R$ and $V_T$ of
stars of the model disk lying in the sector
$|\theta-\theta_\odot|<15^\circ$ along the distance $R$. Shown are
the model dependencies obtained for model 1, which does not take into
account the gravity from the elliptical rings (the red solid line);
for model 2, which takes into account the gravity from three
elliptical rings ($R_1$, $R_2$ and $r$,  the black dashed line); and
for model 3, which takes into account the gravity from only two outer
rings ($R_1$ and $R_2$, the green solid line). The model velocity
profiles were averaged over the time intervals of 0.5 Gyr. The
boundaries of the time intervals are indicated on each frame. It is
clearly seen that taking into account the gravity from the elliptical
rings has little effect on the distributions of the velocities $V_R$
and $V_T$ along the distance $R$. The observational dependences
derived from the {\it Gaia} DR3 data (the blue line)  correspond to
the to the moment  of Gaia observations  (J2016). The vertical lines
show the position of the OLR.} \label{profile}
\end{figure*}
\begin{figure*}
\resizebox{\hsize}{!}{\includegraphics{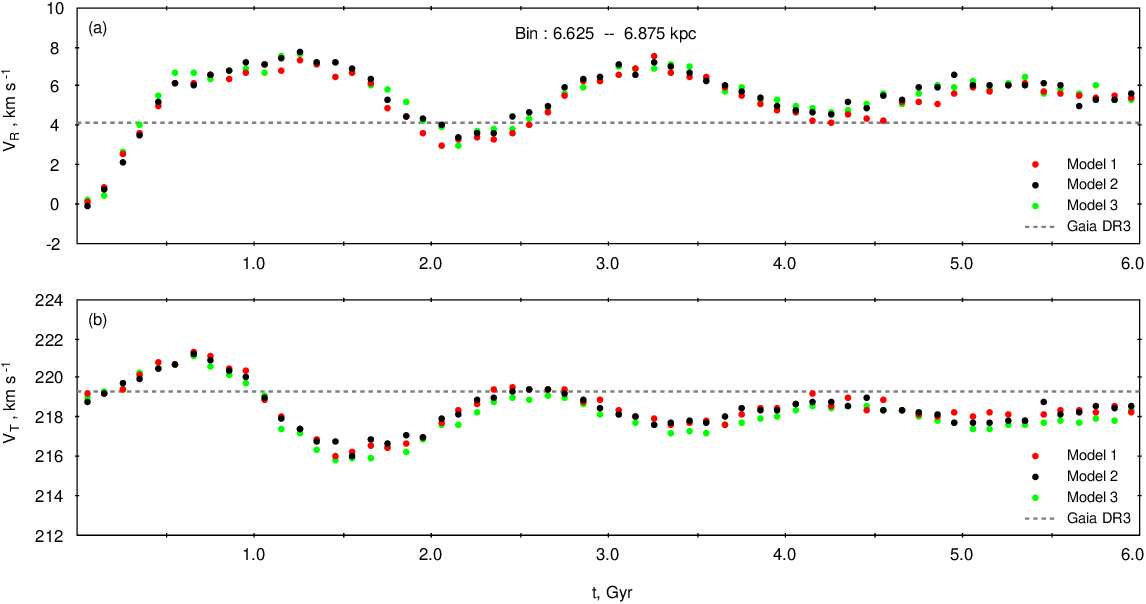}} \caption{Variations
of the median velocities $V_R$ (a) and $V_T$ (b) calculated in the
distance bin $R=6.625$--6.875 kpc as a function of time $t$. Shown
are the velocities computed for model 1, not taking into account the
gravity from the elliptical rings (red circles); the velocities
obtained for model 2, taking into account the gravity from three
elliptical rings ($R_1$, $R_2$, and $r$, black circles); and the
velocities calculated for model 3, taking into account the gravity
from only two outer rings ($R_1$ and $R_2$, green circles). Random
errors in determination of the  median velocities $V_R$ and $V_T$ are
smaller than the size of the circles. The observational velocities
derived from the {\it Gaia} DR3 data correspond to the to the moment
of Gaia observations  (J2016) (the dashed lines). It is clearly seen
that the velocities $V_R$ and $V_T$ in three models oscillate almost
synchronously.} \label{vr_vt_bin}
\end{figure*}
\begin{table*}  \centering \caption{\normalsize Parameters of oscillations of the velocities $V_R$ and $V_T$}
\begin{tabular}{cccc}
\\[-7pt]\hline\\[-7pt]
\multicolumn{4}{c}{$V_R$}\\
Parameters & Model 1 & Model 2 &  Model 3 \\
\\[-7pt]\hline\\[-7pt]
$P$, Gyr & $2.1\pm0.1$ & $2.1\pm0.1$ &$2.1\pm0.1$   \\
$A$, km s$^{-1}$  & $1.76\pm0.15$ & $1.70\pm0.17$ & $1.63\pm0.17$  \\
$\varphi$, deg. & $257\pm5^\circ$ & $247\pm6^\circ$ & $251\pm6^\circ$   \\
$\overline{V_R}$, km s$^{-1}$  & $5.2\pm0.1$ & $5.5\pm0.1$ & $5.4\pm0.1$ \\
\\[-7pt]\hline\\[-7pt]
\multicolumn{4}{c}{$V_T$}\\
Parameters & Model 1 & Model 2 &  Model 3 \\
\\[-7pt]\hline\\[-7pt]
$P$, Gyr & $1.9\pm0.1$ & $1.9\pm0.1$ &$1.9\pm0.1$   \\
$A$, km s$^{-1}$  & $1.24\pm0.14$ & $1.16\pm0.13$ & $1.29\pm0.14$  \\
$\varphi$, deg. & $329\pm6^\circ$ & $327\pm6^\circ$ & $335\pm6^\circ$   \\
$\overline{V_T}$, km s$^{-1}$  & $218.5\pm0.1$ & $218.5\pm0.1$ & $218.2\pm0.1$ \\
\\[-7pt]\hline\\[-7pt]
\end{tabular}
\label{tab_3}
\end{table*}

\subsection{4.3 Distribution of the velocities $V_R$ and $V_T$ along the distance $R$}

We compare the distributions of the radial, $V_R$, and azimuthal,
$V_T$, components of the stellar velocities calculated with and
without taking into account the gravity from the elliptical rings.
Figure~\ref{profile} shows the distributions of the median velocities
of the model-disk stars lying in the $|\theta-\theta_\odot|<15^\circ$
sector along the distance $R$. The median velocities are calculated
in the $\Delta R=250$-pc wide bins. The model dependencies are
obtained for model 1, which does not take into account the gravity
from the elliptical rings (red solid line),  for model 2, which takes
into account the gravity from three elliptical rings ($R_1$, $R_2$,
and $r$, black dashed line), and  for model 3, which takes into
account the gravity from only two outer rings ($R_1$ and $R_2$, green
solid line). Model velocity profiles are averaged over the 0.5-Gyr
time intervals. Also shown are the observational distributions
derived from the {\it Gaia} DR3 data (blue line).

Figure~\ref{profile} shows that the model profiles of the $V_R$ and
$V_T$ velocity distributions over the distance $R$ obtained with and
without  the gravity from the elliptical rings practically coincide.
We consider the $V_R$ and $V_T$ velocity distributions of the model
disk averaged over 5 time intervals (0.5--1.0, 1.0--1.5, 1.5--2.0,
2.0--2.5, 2.5--3.0 Gyr). The standard deviation of the profiles of
the $V_R$-velocity distributions  obtained for models 1 and 2 and
averaged over 5 time periods and 13 distance bins in the
$R=5.875$--9.125-kpc range is 0.31 km s$^{-1}$. For the azimuthal
velocity profiles, the similar estimate is 0.18 km s$^{-1}$. The
standard deviation of the profiles of the $V_R$ ($V_T$) velocity
distributions  obtained for  model 1 and 3 is 0.30 (0.26) km
s$^{-1}$.

Note that the model and observational profiles differ much more. The
standard deviation of the profiles of the  distributions of the
velocity $V_R$ ($V_T$) calculated for model 1 and the velocity $V_R$
($V_T$) derived from the {\it Gaia} DR3 data is 1.19 (1.45) km
s$^{-1}$.

As for random errors, the median velocities of  stars in the model
disk calculated in the  $\Delta R=250$-pc wide bins in the range
$R\approx 6$--9 kpc and averaged over the time intervals of 0.5 Gyr
are determined with an error of less than $0.1$ km s$^{-1}$,
therefore the standard deviation of 0.31 km s$^{-1}$, obtained for
models 1 and 2, cannot be random at the significance level of $P \sim
2\sigma$. The median velocities calculated for the observational
distribution in the distance range $R \approx 6$--9 kpc also do not
exceed $\sim 0.1$ km s$^{-1}$, therefore the deviation of 1.19 km
s$^{-1}$ is significant at the level of $P \sim 8\sigma$. Here we use
the hypothesis of a normal distribution of stellar velocities in bins
relative to the median values.

So far we have studied the deviations of the median velocities
averaged in 13 bins over distance. Now let us consider one distance
bin, $R=6.625$--6.875 kpc, in which the amplitude of the velocity
variation $V_R$ has  maximum value \citep{melnik2024}.

Figure~\ref{vr_vt_bin} shows the median velocities $V_R$ (a) and
$V_T$ (b) calculated in the bin $R=6.625$--6.875 kpc for  three
models as a function of  time, $t$.  In this bin, the observed
velocities derived from the {\it Gaia} DR3 data have the values of
$V_R=4.15$ and $V_T=219.30$ km s$^{-1}$ (gray dashed lines). The
model velocities $V_R$ and $V_T$ are averaged over the time intervals
of 100 Myr (10 values, spaced by 10 Myr), and their random errors
equal 0.14 and 0.10 km s$^{-1}$, respectively. The velocities
calculated for models 1, 2 and 3  are shown in red, black, and green,
respectively. It is clearly seen that the oscillations of the
velocities $V_R$ and $V_T$ calculated for three models occur almost
synchronously.

To quantitatively confirm the fact of the synchronous oscillations,
we determine the parameters of  oscillations of the velocities $V_R$
and $V_T$ in the bin $R=6.625$--6.875 kpc for  three models
(Fig.~\ref{vr_vt_bin}). Table~3 presents the period, $P$, amplitude,
$A$, initial phase of oscillations, $\varphi$, and the average
velocity, $\overline{V_R}$ ($\overline{V_T}$), obtained for the
velocities $V_R$ (upper block) and $V_T$ (lower block). The method of
determination of these parameters is described in detail in
\citet[][section 3]{melnik2024}. It is clearly seen that the values
of the parameters $P$, $A$, and $\varphi$ obtained for different
models coincide within the errors, but the values of the average
velocity $\overline{V_R}$ ($\overline{V_T}$) calculated for different
models differ at the significance level of $\sim 2\sigma$. However,
it is the parameters $P$ and $\varphi$ that determine the positions
of the maxima and minima in the velocity oscillations. Therefore, the
oscillations of the $V_R$ ($V_T$) velocity in three models occur
synchronously within the errors.

Figure~\ref{vr_vt_bin} allows us to identify the time periods when
the model (circles) and observational (dashed lines)  velocities
calculated for the bin $R=6.625$--6.875 kpc coincide within the
errors. For the $V_R$ velocity, these time periods lie near the
moments $t=1.8$, 2.5, and 4.2 Gyr from the start of simulation. We do
not consider the moment 0.3 Gyr, since the bar has not yet formed by
this time ($T_g=0.45$ Gyr). For the $V_T$ velocity, we can identify
intervals near the moments $t=1.0$, 2.5, and 4.2 Gyr. In general,
these results agree well with the age estimates of the Galactic bar,
$2.5\pm0.3$ or $4.5\pm0.5$ Gyr, corresponding to the best agreement
between the model and observational velocities $V_R$ obtained for 13
bins in the  range $\sim 6$--9 kpc \citep[more,][section
10]{melnik2024}.

Thus, the models constructed with and without taking into account the
gravity from the elliptical rings reproduce equally well  the
observational  distributions of the velocities $V_R$ and $V_T$ along
the distance $R$. Generally, this result is expected, since the
masses of the rings are small and the gravitational perturbations
from them must have little effect on the process of tuning the
epicyclic motions of stars near the OLR of the bar.

\section{5. CONCLUSIONS}

We investigated the influence of gravitational forces generated by
the elliptical resonance rings on the kinematics of old-disk stars
near the OLR of the bar. The model disk forms the inner ring, $r$,
located near the CR of the bar, and outer rings, $R_1$ and $R_2$,
located near the OLR of the bar. We used three models of the Galaxy:
model 1 does not take into account the gravity from the elliptical
rings \citep[for more details, see][]{melnik2021}, model 2 includes
gravitational forces from three elliptical rings ($R_1$, $R_2$, and
$r$) and model 3 takes into account the gravity from only two outer
elliptical rings ($R_1$ and $R_2$).

Using the program that enhances the contrast, we determined the
positions of the rings in the model disk. The boundaries and the
midline of each ring were represented by similar ellipses. The
average density of stars $\Delta \Sigma$ in the rings was calculated
relative to the density at the time moment $t=0$, when the
distribution of stars in the model disk was exponential
(Subsection~4.1).

The density $\Delta \Sigma$ in the outer rings $R_1$ and $R_2$
oscillate practically in antiphase with the period of $P=2.0\pm0.1$
Gyr (Fig.~\ref{oscil}a), which is close to the median value of the
period of oscillations of  librating orbits \citep{melnik2024}. In
our calculation of gravitational forces, we did not take into account
the fluctuations in the ring density. To estimate the masses of the
rings $R_1$ and $R_2$, we used  maximum values of the density $\Delta
\Sigma$ after the bar reaches its full power, $t>0.45$ Gyr (Table~1).

The density perturbation $\Delta \Sigma$ in the inner ring, $r$,
oscillates near the average value with the period of $P=0.58\pm0.01$
Gyr (Fig.~\ref{oscil}b), which is close to the period of oscillations
of stars in long-period orbits near the equilibrium points $L_4$ and
$L_5$. For the inner ring, we used the average density $\Delta
\overline{\Sigma}$ after the bar reaches its full power (Table~1).

We suppose the gravitational potential of the Galactic disk  to be a
superposition of the potentials of the exponential disk and the
elliptical rings. The stellar density  across an elliptical ring is
assumed to vary in a cosine law (Eq.~\ref{cos}). Using the numerical
differentiation of the potential, we calculated the radial, $F_R$,
and azimuthal, $F_T$, components of the gravitational force acting on
a particle of unit mass from the elliptical ring. The force $F_R$ is
always directed toward the midline of the ring: in the inner region
of the ring it is directed away from the center of the Galaxy while
in the outer region it is directed toward the center. At the  midline
of the ring the force $F_R$ is zero (Subsection 4.2).

We represented the distribution of the force $F_R$ along the distance
$R$ as a combination of three polynomials in powers of $R/R_e$ or
$R_e/R$, where $R_e$ is the distance from the center of the Galaxy to
the midline of the ring for a given  angle $\theta$. If the point
under consideration lies inside the inner boundary of the ring, i.~e.
$0<R<R_{in}$, then $F_R$ can be represented by a fifth-order
polynomial in powers of $p=R/R_e$ (Eq.~\ref{fr_a}); if the point lies
inside an elliptic ring, i.~e. $R_{in}\le R \le R_{out}$, then $F_R$
can be calculated using a fourth-order polynomial in powers of
$s=R_e/R$ (Eq.~\ref{fr_b}); and finally, if the point lies outside
the outer boundary of the ring, $R > R_{out}$, then $F_R$ can be
calculated using another fourth-order polynomial in powers of
$s=R_e/R$ (Eq.~\ref{fr_c}).

The similarity in the profiles of the distributions of the forces
$F_R$ and $F_T$ along the distance $R$ allows us to calculate the
force $F_T$ through the force $F_R$ (Eq.~\ref{ft_d}).

The forces $F_R$ ($F_T$) calculated using numerical differentiation
(Eq.~\ref{d_fr} and \ref{d_ft}) and using analytical representation
(Eq.~\ref{fr_a}--\ref{ft_d}) for the rings $r$, $R_1$ and $R_2$
differ by no more than 5.7, 5.5 and 4.3\% (1.3, 1.2 and 1.0\%) from
maximum value of the force $F_R$ created by the elliptical ring,
respectively (Fig.~\ref{force}).

We built the distributions of the radial $V_R$ and azimuthal $V_T$
velocities of stars of the model disk  lying in the sector
$|\theta-\theta_\odot|<15^\circ$ along the distance $R$. The model
dependencies were obtained with and without taking into account the
gravity from the rings $R_1$, $R_2$ and $r$ (Fig.~\ref{profile}). The
median velocities $V_R$ and $V_T$ were calculated in the 250-pc wide
distance bins and were averaged over the time intervals of 0.5 Gyr.
The standard deviation of the $V_R$ ($V_T$) velocity profiles
obtained with and without  the gravity from the three elliptical
rings ($R_1$, $R_2$, and $r$) in the distance range $R \approx 6$--9
kpc is 0.31 (0.18) km s$^{-1}$. This is noticeably smaller than the
standard deviation of the model (model 1) and observational profiles
of the distribution of the $V_R$ ($V_T$) velocity equal to 1.19
(1.45) km s$^{-1}$. The random errors in determination of the median
velocities  in bins in the range $R \approx 6$--9 kpc for  the model
and observational data do not exceed $\sim 0.1$ km s$^{-1}$
(Subsection 4.3).

We selected one distance bin, $R=6.625$--6.875 kpc, corresponding to
the largest amplitude of oscillations of the velocities $V_R$ and
$V_T$ \citep{melnik2024} and studied the oscillations of the
velocities calculated for three models (Fig.~\ref{vr_vt_bin}). It
turned out that oscillations of the velocities $V_R$ and $V_T$
calculated with and without the gravity from the elliptical rings
occur synchronously within the errors (Fig.~\ref{vr_vt_bin},
Table~3).

In general, the gravity of the elliptical rings has little effect on
the process of tuning  epicyclic motions near the OLR of the bar.

\section*{ACKNOWLEDGMENTS}

We thank the anonymous reviewer for  interesting discussion and
helpful comments. This work was carried out using the data from the
European Space Agency (ESA) mission Gaia
(https://www.cosmos.esa.int/gaia), processed by the Data Processing
and Analysis Consortium (DPAC,
https://www.cosmos.esa.int/web/gaia/dpac/consortium) Gaia. DPAC
support was provided by national institutions, in particular,
institutions participating in the multilateral agreement Gaia.

\section*{FUNDING}

The study was conducted under the state assignment of Lomonosov
Moscow State University. E.N. Podzolkova is the recipient of a
scholarship from the Foundation for the Development of Theoretical
Physics and Mathematics "BASIS" (grant No. 21-2-2-44-1).

\section*{CONFLICT OF INTEREST}

The authors of this work declare that they have no conflicts of
interest.

\end{document}